# Autonomous atomic Hamiltonian construction and active sampling of x-ray absorption spectroscopy by adversarial Bayesian optimization

Yixuan Zhang[1], Ruiwen Xie[1], Teng Long[1,2], Damian Günzing[3], Heiko Wende[3], Katharina J. Ollefs[3], Hongbin Zhang[1,*]

1. Institute of Materials Science, Technical University of Darmstadt, Darmstadt 64287, Germany
2. School of Materials Science and Engineering, Shandong University, Jinan 250061, China
3. Faculty of Physics, University of Duisburg-Essen, Duisburg 47057, Germany

*Corresponding to Hongbin Zhang, hongbin.zhang@tu-darmstadt.de

## Abstract

X-ray absorption spectroscopy (XAS) is a well-established method for in-depth characterization of the electronic structure due to its sensitivity to the local coordination and electronic states of the active ions. In practice hundreds of energy points should be sampled during the XAS measurement, most of which are redundant and do not contain important information. In addition, it is also a tedious procedure to estimate reasonable parameters in the atomic Hamiltonian for mechanistic understanding. We implemented an Adversarial Bayesian optimization (ABO) algorithm comprising two coupled BOs to automatically fit the multiplet model Hamiltonian and meanwhile to sample effectively based on active learning. Taking NiO as an example, for simulated spectra which can be well fitted by the atomic model, we found that less than 30 sampling points are enough to obtain the complete XAS with the corresponding crystal field or charge transfer model, which can be selected based on intuitive hypothesis learning. Further application on the experimental spectra, it revealed that less than 80 sampling points can already give reasonable XAS and reliable atomic model parameters. Our ABO algorithm has a great potential for future application in automated physics-driven XAS analysis and active learning sampling.

## Introduction

XAS is an important experimental technique used in the investigation of material properties, where a large number of sampling points are required to be measured to capture the fine details and the experimental data need to be analyzed based on atomic Hamiltonian with undetermined parameters[1,2]. Various methods have been used to interpret XAS, including density functional theory (DFT)[3,4], semi-empirical multiplet models[5,6], and DFT combined with dynamical mean-field theory (DFT + DMFT)[7]. The DFT-based methods are capable of capturing the crystalline environment of the active ions but limited to the complications arising from the correlation effects. The DFT + DMFT calculations simultaneously account for the chemical realism and correlation effects, where the spectroscopy is evaluated based on the Anderson impurity model (AIM)[8]. Unfortunately, the DFT + DMFT calculations require a large amount of computational time and resources, thus are usually performed off-site by experts.

In contrast, the semi-empirical methods based on atomic Hamiltonians consider the essential parameterized interactions such as Coulomb interactions, spin-orbit interactions, crystal field, charge transfer and broadening, resulting in transparent mechanistic understanding of XAS [9–11]. For instance, the crystal field multiplet (CFM) theory and charge transfer multiplet (CTM) theory constitute two popular models in understanding and simulating experimental XAS of transition metal (TM) oxides. CFM describes atomic interactions with the surrounding ligands treated as a perturbation using the effective electric field[12], however, the neglection of covalency makes it challenging to describe the crystal field splitting. In contrast, the CTM model takes metal-ligand charge transfer into consideration. Sugano and Shulman devised molecular orbitals (MOs) model to calculate the ligan-field parameters and got good consistency with experimental results[13]. Moreover, Fujimori and Minami used the cluster model considering strong 3d-3d correlations and 3d-ligand orbitals hybridization to improve the interpretability of MLFT for the valence-band photoemission[14,15]. In order to extrapolate the MLFT further to understand band gaps and broad satellite in XAS of transition metals, Gunnarsson, van der Laan and Zaanen took the large ligand hole bandwidth into the account by employing the Anderson impurity Hamiltonian[16–19].

Another practical challenge is how to perform efficient sampling during the experimental measurements. A homogeneous energy grid with several hundreds of measurements within specific energy range is often needed, which is inefficient and resource-costly. The quality of the spectrum varies depending on the choice of sampling points. One way out is the so-called design of experiments (DoE)[20], which focuses on selecting the input variables of experiments with a significant effect on experiments' target value, and examining how to choose the best combination of independent variables. However, a good number of sampling points are still needed in order to obtain a spectrum with a sufficient signal-to-noise ratio. With the help of machine learning (ML), DoE has undergone a transformation. Several ML-based methods have been applied to the prediction and design of spectroscopy experiments[21–23]. Nevertheless, collecting the needed data for ML still requires a lot of effort. The active learning (AL) method based on Bayesian analysis can be applied to reduce the required data, where the posterior mean and variance of the parameter space can be calculated using the Gaussian process (GP) based on the prior distribution function[24]. Mostly, AL stops once the desired level of performance is achieved[25], which can be applied to spectroscopic experiments to design appropriate stopping criteria. For instance, the decision to stop the measurements can be made by estimating the physical parameters to determine the validity of the current experimental results[26,27]. However, such parameter assessment often requires comparisons between the experimental and theoretical spectra, which are usually not available during the experiments. Furthermore, automatic stopping criteria based on generalization error analysis have now been applied to spectral AL to circumvent this complex parametric analysis process[23]. While such a combination of active learning and stopping criterion can effectively reduce the number of sampling points, the absence of supports based on theoretical models reduces the reliability of AL to extract points of important features. Hence, there is a strong impetus to develop a physics-driven stopping criterion for experimental sampling.

In this work, we proposed an ABO algorithm, comprising automated parameter fitting to obtain the atomic Hamiltonian and efficient sampling based active-learning. Taking NiO as an example with theoretical XAS obtained using Quanty[9], we demonstrated that both CFM and CTM models with up to 15 parameters can be constructed, while keeping the number of sampling points at the order of 30. To identify the dominant physical processes, hypothesis learning was done which can distinguish the CFM and CTM models depending on the standard real spectrum (The ground truth in ML). It was further revealed that our ABO algorithm can be applied on the real experimental data, providing a valuable solution to the currently time-consuming XAS measurements and analysis.

**Methods**

**Multiplet Models**

For NiO, the $L_{2,3}$ edge XAS of $Ni^{2+}$ can be evaluated using Quanty with model Hamiltonians based on either CFM or CTM defined as follows,

$$H_{CFM} = H_U^{dd} + H_U^{pd} + H_{l \cdot s}^d + H_{l \cdot s}^p + H_{CF} + H_{ex}, \quad (1)$$

$$H_{CTM} = H_U^{dd} + H_U^{pd} + H_{l \cdot s}^d + H_{l \cdot s}^p + H_o^p + H_o^d + H_o^L + H_{hyb}^{dL} + H_{ex}, \quad (2)$$

with $H_U^{dd}$ being the on-site Coulomb repulsion between $3d$ electrons of Ni, $H_U^{pd}$ being the on-site Coulomb interaction between Ni $2p$ and $3d$ electrons, $H_{l \cdot s}^d$ and $H_{l \cdot s}^p$ being the spin-orbital coupling terms corresponding to Ni $3d$ and $2p$ orbitals, respectively, $H_{ex}$ being the Weiss magnetic field acting on Ni. To understand the underlying physics, such empirical parameters should be tailored so that the experimental XAS can be well reproduced, which is challenging as it is a multi-dimensional fitting problem (cf. Supplementary for the complete lists of all relevant atomic parameters). For instance, CFM and CTM distinguish from each other by the excited charge dynamics, where CFM allows only excitations based on the local crystal fields (denoted by $H_{CF}$ in Eq.(1)), whereas virtual charge transfer processes are considered between Ni-$3d$ and ligand $2p$ orbitals in CTM model (represented by $H_{hyb}^{dL}$ in Eq.(2)). In the calculations of XAS using Quanty, the two Hamiltonians defined in Eqs.(1) and (2) were applied for both the ground states and excited states. After calculating the ground state wave function, XAS was evaluated using a Lanczos-based Green's function method[28,29]:

$$G(\omega) = \left\langle \Psi T^\dagger \left| \frac{1}{\omega + i\frac{\Gamma}{2} - H_{CFM/CTM}^{fs}} \right| T\Psi \right\rangle, \quad (3)$$

where $|\Psi\rangle$ denotes the ground state, $T$ is the transition operator (here it represents a $2p3d$ dipole excitation), $H_{CFM/CTM}^{fs}$ is the final state Hamiltonian, $\omega$ is the energy relative to the energy of state $|\Psi\rangle$, and $\Gamma$ denotes the core-hole lifetime. For the specific forms of each term in Eqs. (1), (2) and (3), please refer to the Supplementary Information. In order to acquire better fitting the experimental data, the Gaussian broadening is usually added uniformly to mimic the broadening caused by experimental instrument while the energy-dependent Lorentzian broadening is applied to simulate the core-hole lifetime broadening.

**Adversarial Bayeian Optimization algorithm**

Following the idea of Generative Adversarial network (GAN)[30] where both the generators and discriminators are trained simultaneously, we combined the multiplet Hamiltonian construction and active learning sampling together, leading to an ABO algorithm. Similar to GAN, where a trained generator minimizes the difference between the existing and generated data and the discriminator aims to maximize it, the multiplet model fitting acts like the generator to minimize the difference between the model predicted XAS and the real XAS, whereas the active learning sampling behaves like the discriminator to identify the next sampling points with large uncertainties measured by the loss function. Correspondingly, in our ABO algorithm, there are two coupled BO algorithms. The first one, called fitting BO (fBO), is a trust region Bayesian optimization (TuRBO) applied on our multiplet Hamiltonian to search for better parameters minimizing the difference between the theoretically predicted XAS and the real XAS values. The second one, called sampling BO (sBO), is a simple BO, which functions to select the sampling points that maximize the differences between the predicted and real XAS values. The mathematical form of ABO is expressed as:

$$\max_{sBO}(\min_{fBO}(\frac{\sum|Y_{i,experiment} - Y_{i,Quanty}|}{N})), \qquad (4)$$

where $Y_{i,\text{Quanty}}$ denotes the theoretically predicted XAS obtained using the Quanty code starting from the multiplet Hamiltonian, and $Y_{i,experiment}$ indicates the real XAS obtained either from Quanty simulations (with hidden parameters) or real experimental measurements, and $N$ is the size of current samples set.

For fBO, we cannot ignore the possibility that we still may end up with a local optimum due to the high dimensionality of the fitting problem. In order to ensure that the fitting is accurate enough, we set a criterion that the loss between the predictions and the truth should be smaller than a threshold TH, which is defined as the average deviation between all the experiment values and its corresponding predictions. Afterwards, the resulting parameters and XAS will be fed into sBO, which will evaluate the current points and their corresponding differences, and then calculate the location of next point which is thought to have maximum difference. The ABO workflow is shown in Figure 1.

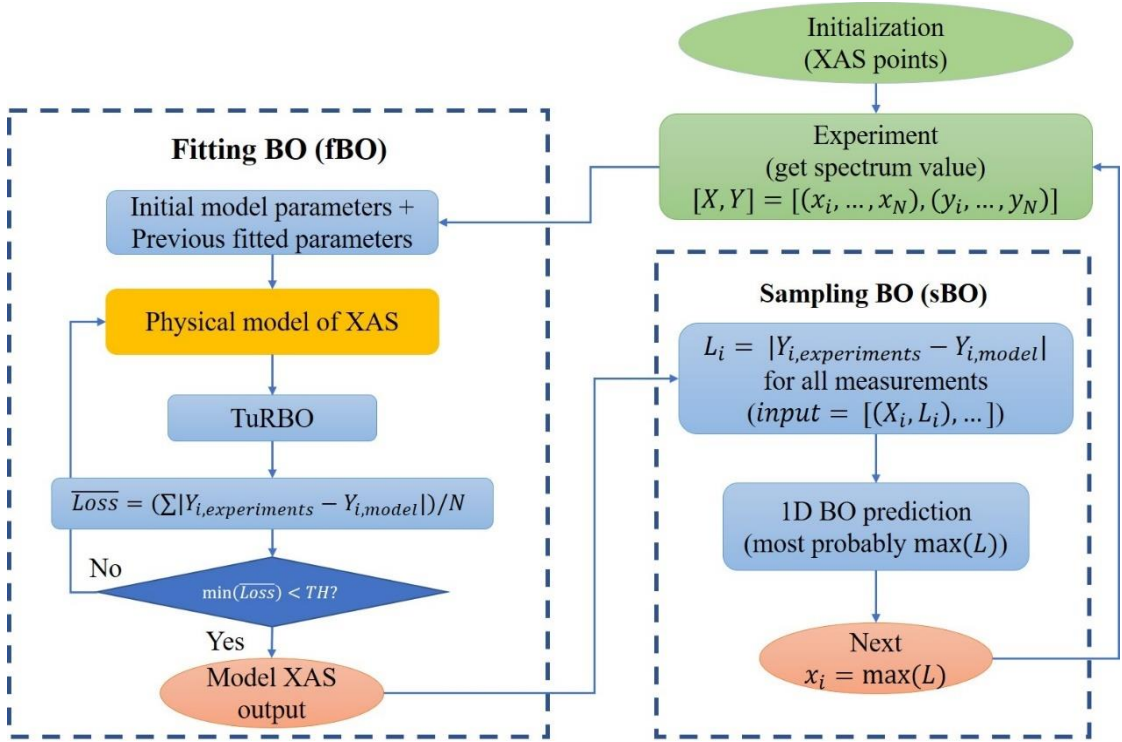

Figure 1. The workflow of adversarial Bayesian optimization (ABO) algorithm

**Bayesian optimization**

Inside the ABO algorithm, we used an ExactGP and the scaled Matern52 kernel implemented in Gpytorch[31] to interpolate the intensity values of sampling points of XAS by considering similarity of points and estimating the value of functions correspondingly[24]. Both fBO and sBO were constructed by using BoTorch[32]. We used GP to describe the distribution of XAS measurements. The outcomes of sampling points were first normalized to zero mean and unit variance. Assuming the observed inputs are $X_n = (x_1, ..., x_n)^T \in \mathbb{R}^n$ and the corresponding outcomes are $Y_n = (y_1, ..., y_n)^T$. The relation between the inputs and the outcomes follows the rule as $y_i = f(x_i) + \sigma_i$, where $\sigma_i$ is a Gaussian which describes the noise distribution. The Gaussian of the previous observation and the new point can be expressed as:

$$Y_n \sim GP(0, K(X_n, X_n) + \sigma_n^2 I), \quad (5)$$

where the $K(X_n, X_n) + \sigma_n^2 I$ is the covariance matrix between all observations with $\sigma_n$ the estimated noise distribution during the measurement, $I$ is an identity matrix. The kernel function used is Matern52 with a characteristic length-scale $l$. The twice differentiability of Matern52 makes it preferred to mimic the noise behavior in ML. However, the smoothness of the simulated function varies with respect to different dimension and location of the parameters. Thus, with a uniform $l$ in the whole parameter space, the changes of local smoothness cannot be well described. In Gpytorch, the magnitude of the kernel is scaled by a scale factor $\theta_{scale}$. As for the position of the parameter space, the $l$ will be adjusted according to samples in batch and the range of Trust Region (TR) in BoTorch. This part will be explained later in the Trust Region Bayesian Optimization section. Since GP is closed in the conditional action and marginalization cases, length scale $l$ can be optimized by maximizing the log marginal likelihood:

$$\underset{l}{argmax}(\log(L(X_n, Y_n; l))), \qquad (7)$$

where:

$$\log(L(X_n, Y_n; l)) = -\frac{1}{2}\log|K(l) + \sigma_n^2 I| - \frac{n}{2}\log(2\pi) - \frac{1}{2}Y_n^T(K(l) + \sigma_n^2 I)^{-1}Y_n$$

This regression is done by Adam optimization algorithm implemented in BoTorch.

For the next energy point $x^*$, its posterior distribution is also a Gaussian:

$$y^*|Y_n \sim GP(\bar{\mu}(x^*), \bar{K}(x^*, x^*)), \qquad (8)$$

with

$$\bar{\mu}(x^*) = K(X_n, x^*)^T (K(X_n, X_n) + \sigma_n^2 I)^{-1} Y_n, \qquad (9)$$

$$\bar{K}(x^*, x^*) = K(x^*, x^*) - K(X_n, x^*)^T (K(X_n, X_n) + \sigma_n^2 I)^{-1} K(X_n, x^*), \qquad (10)$$

Here $y^*$ is the measured value of point $x^*$, $\bar{\mu}(x^*)$ is the posterior mean and $\bar{K}(x^*, x^*)$ the posterior variance of point $x^*$, $K(X_n, x^*) = (K(x_1, x^*), \ldots, K(x_n, x^*))$ is the vector of covariance between $x^*$ and all observations, which only varies with the position of $x^*$ and the noise estimation.

After obtaining the posteriors, we can use them to find the next sampling point with the help of acquisition function, which balances between exploration and exploitation. We used the Upper Confidence Bound (UCB) and Parallel Upper Confidence Bound (qUCB)[33] implemented in BoTorch as the acquisition function for sBO and fBO (with Thompson sampling), respectively. The hyperparameters used for UCB and qUCB were adjusted accordingly in the different tests. For the CFM model, the global hyperparameters of sBO were set to 16; for the CTM model the hyperparameters were 49 for the initial 10 iterations and 4 for the remaining steps. Whereas for the tests using the experimental data, the hyperparameters were set to 100 for the initial 18 iterations and 1 for the remaining steps. For all simulations, the reason for using high values at the beginning is that our ABO starts with a small initial sample size, such high exploration values can give us a rough overall data distribution and prevent the ABO from ending at a local minimum.

**Trust Region Bayesian Optimization**

Although the traditional Bayesian optimization shows its power in low-dimensional gradient-free fitting, it is still limited by the exponential growth in computational complexity with high-dimensional parameters and large sample size[34–36]. For the high-dimensional black-box problem like CFM and CTM model, it is difficult to describe this complex distribution precisely with a GP with a stationary kernel. In addition, the Bayesian optimization tends to stop at a local optimum due to the non-convexity of the problem[37]. To address these challenges, we applied the trust region (TR) method into the optimization. In the TR method, the algorithm selects the candidates according to each different TR, introduces a separate GP surrogate model for each TR, and optimizes its hyperparameters accordingly. Thus, within TR, the surrogate model is sufficiently accurate to describe the problem locally and is able to provide an optimal solution[38]. During the optimization, TR is chosen to be a hyper-rectangle area with the current best solution $x^*$ located at the center of it. The base side length of each dimension of this hyper-rectangle is set to be $L$, and the actual side length of each dimension of parameters can be obtained by $L_i = \lambda_i L / \left(\prod_{j=1}^{d} \lambda_j\right)^{\frac{1}{d}}$, where $\lambda_i$ is the length scale of parameters in GP. In each step we selected a batch of candidates within the TR with the help of GP and acquisition function. If the algorithm finds a better optimal solution than the current one in two successive iterations, the current TR is treated to be small enough so that the existing GP surrogate model is already capable of predicting with good accuracy. The $L$ of TR is then doubled to allow the surrogate model to search in a larger range. If the algorithm cannot find a better solution in two iterations, the current TR is considered to be too large and will be halved. The algorithm will stop when the side length $L$ meets maximum threshold $L_{\max}$ or minimum threshold $L_{\min}$. It is noted that a single TR cannot guarantee a global optimum but mostly the local minimum. To solve this problem, a parallelization of TRs with varying local GPs was introduced by TuRBO. The best result is then selected by:

$$x_i^{(t)} \in \underset{l}{\arg\min} \underset{x \in TR_l}{\arg\min} f_l^{(i)}, \qquad (11)$$

where $x_i^{(t)}$ the optimal point of our selection; $l \in \{1, 2, \ldots, m\}$ denotes all the TRs utilized in each iteration, $f_l^{(i)} \sim GP_l^{(t)}(\mu_l(x), k_l(x, x'))$ is the GP surrogate model applied in $TR_l$[39].

**Results**

**Fitting CFM with active learning sampling**

Keeping in mind that the goal of the ABO algorithm is to reach the standard real spectrum with as few sampling points as possible. Starting with CFM fitting, there are 9 independent parameters, including the Coulomb interaction $F_{dd}^2$, $F_{dd}^4$, $F_{pd}^2$, $G_{pd}^1$, $G_{pd}^3$, the crystal splitting $10D_q$, the spin orbital coupling strength of Ni $2p$ and $3d$ orbitals $\xi_{2p}$, $\xi_{3d}$, the exchange magnetic field $B_{ex}$. The Slater integral $F_{dd}^0$ is related to $U$ by

$$F_{dd}^0 = U_{dd} + \frac{2}{63}(F_{dd}^2 + F_{dd}^4), \qquad (12)$$

and $F_{pd}^0$ is related to $U_{pd}$ by

$$F_{pd}^0 = U_{pd} + \frac{1}{15}G_{pd}^1 + \frac{3}{70}G_{pd}^3. \qquad (13)$$

Note that as the total number of electrons is conserved in CFM, $U$ and $U_{pd}$ can be dropped out, *i.e.*, $U = U_{pd} = 0$.

To demonstrate the efficacy and robustness of the ABO algorithm, it is first applied to the Ni$^{2+}$ L$_{2,3}$ edge XAS curve produced by the 9-parameter CFM. We note that the L$_{2,3}$ XAS spectra have been already well reproduced using crystal field based models for NiO[40]. In order to reduce the dimension of the problem, $F_{dd}^4$ is fixed based on the feature importance (FI) analyze for all 9 parameters (cf. Supplementary Figure S1). Figure 2 shows the snapshots obtained based on the ABO algorithm with 3, 19, and 27 sampling points (Figure 2.a to 2.c). For an initial dataset consisting of 3 random points (Figure 2.a), it is observed that the resulting XAS (red curve) obtained using the Hamiltonian from fBO has a huge difference compared to the theoretical standard real

XAS curve (black curve). This suggests that due to insufficient sampling points, the Hamiltonian fitting suffers from an extremely non-convex problem, and the resulting CFM model is trapped in one of many possible local minima. However, such an inaccurate Hamiltonian can still be applied to predict the next sampling point in sBO by evaluating the deviation from the theoretical standard real XAS curve. In follow-up iterations, CFM gets substantially improved and the non-convexity is tremendously reduced, reaching the theoretical standard real XAS curve expeditiously (Figure 2.d). With 19 sampling points as shown in Figure 2.b, the algorithm is able to recover the shape of the XAS curve and reproduce the fine structures besides peaks and valleys. Further sampling up to 27 points allows the ABO algorithm to well get the theoretical standard real XAS curve, with the corresponding parameters in CFM converged excellently (cf. Table S3).

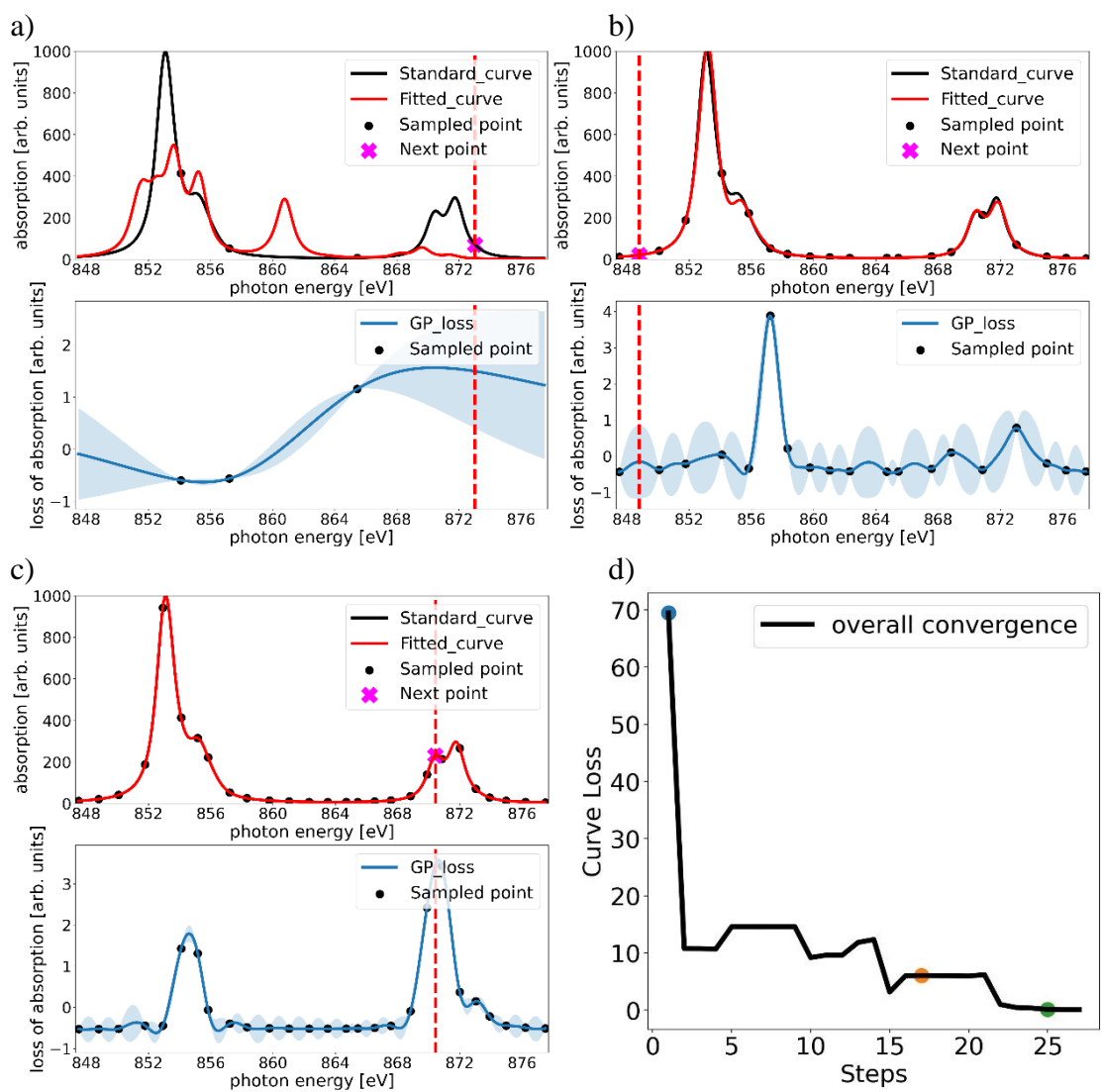

Figure 2. The fitting results of CFM model with: a) 3 sampling points (Step 1); b) 19 sampling points (Step 17); c) 27 sampling points (Step 25); d) CFM XAS curve convergence, with the colored dots denote in order the results of Figure 2.a to 2.c. The upper half plots show the fitted results of fBO. The black curve denotes the standard real values of the unknown XAS curves, the black round dots are the already measured points, the red curve is the fitted result based on current sampled points using fBO

fitting, and the magenta cross is the point to be measured in the next round, which is determined by sBO. In the lower half plots are the standardized prediction of sBO, where the blue dots denote the GP predicted loss with the yellow region are the uncertainties. The parameter comparison for listed steps is shown in Supplementary Table S4.

Figure 2.d shows the convergence indicated by the loss function plotted with respect to the number of sampling points, where the loss function is defined as the average deviation between all points on the prediction and the theoretical standard real XAS curves. Correspondingly, the convergence of model parameters is shown in Figure S3, with the intermediate parameters and the real parameters for the theoretical standard real XAS listed in Table S3. It is observed that $G_{pd}^3$ and $F_{pd}^2$ have relative larger deviation of respectively 2.07% and 1.15% compared to the other parameters, which can be attributed to the less FI for such parameters (see supplementary section S1.1).

**Fitting CTM model with active learning sampling**

To further explore the capability of the ABO algorithm, it is applied to the more sophisticated CTM model, which considers the charge transfer effects. CTM for NiO has in total 15 input parameters, including the Coulomb interaction parameters $U_{dd}$, $U_{pd}$, $F_{dd}^2$, $F_{dd}^4$, $F_{pd}^2$, $G_{pd}^1$, $G_{pd}^3$, the charge transfer energy $\Delta$, the crystal field splittings corresponding to Ni $3d$ and ligand orbitals $10D_q$, $10D_{qL}$, the hopping integral between O-$2p$ and Ni-$3d$ orbitals of $t_{2g}$ and $e_g$ symmetry $V_{t_{2g}}$, $V_{e_g}$, the spin orbital coupling strength of Ni-$2p$ and $3d$ orbitals $\xi_{2p}$, $\xi_{3d}$, and the exchange magnetic field $B_{ex}$. We adapted parts of the algorithm process by giving the physical preconditions to reduce the difficulty of high-dimensional fitting. The adapted ABO algorithm was then straightforwardly applied to fit the XAS curves simulated using the CTM model.

The results are shown in Figure 3. Initial random sampling points are helpful to get a reasonable starting model as we observed for CFM, thus for CTM 10 initial points (Fig. 3.a) are used in order to get a rough estimation of the overall data distribution. This helps significantly with the convergence, in comparison with only one initial point as shown in Fig. S4. And as illustrated in Figure 3.b, 23 sampling points are already enough for the ABO algorithm to capture the essential features such as peaks and satellites. Interestingly, as shown in Figure 3.c, we found that the ABO algorithm can identify the critical regions and keep on improving its accuracy therein, though the fitted curves and used parameter values are already very close to the theoretical standard real spectrum. The convergence of the model parameters is depicted in Fig. S4, and the final fitted parameters are summarized in Table S5. There are four parameters with larger deviations i.e., $V_{t2g}$, $10D_q$, $10D_{qL}$ and $B_{ex}$ (3.01%, 2.83%, 2.24% and 2.11%). According to the FI analysis (see Supplementary Section S1.2), such parameters have relatively less significant FI, thus the sensitivity of the final results with respect to such parameters is marginal. Moreover, the distribution of the selected sampling points covers mostly the regions containing important information on the theoretical standard real spectrum, *e.g.,* the peaks, satellites, and the splitting. This confirms the robustness of our ABO algorithm, *i.e.*, the adversarial algorithm can dynamically allocate the sampling region to improve the XAS while continuously improving the parameter fitting for the atomic Hamiltonian.

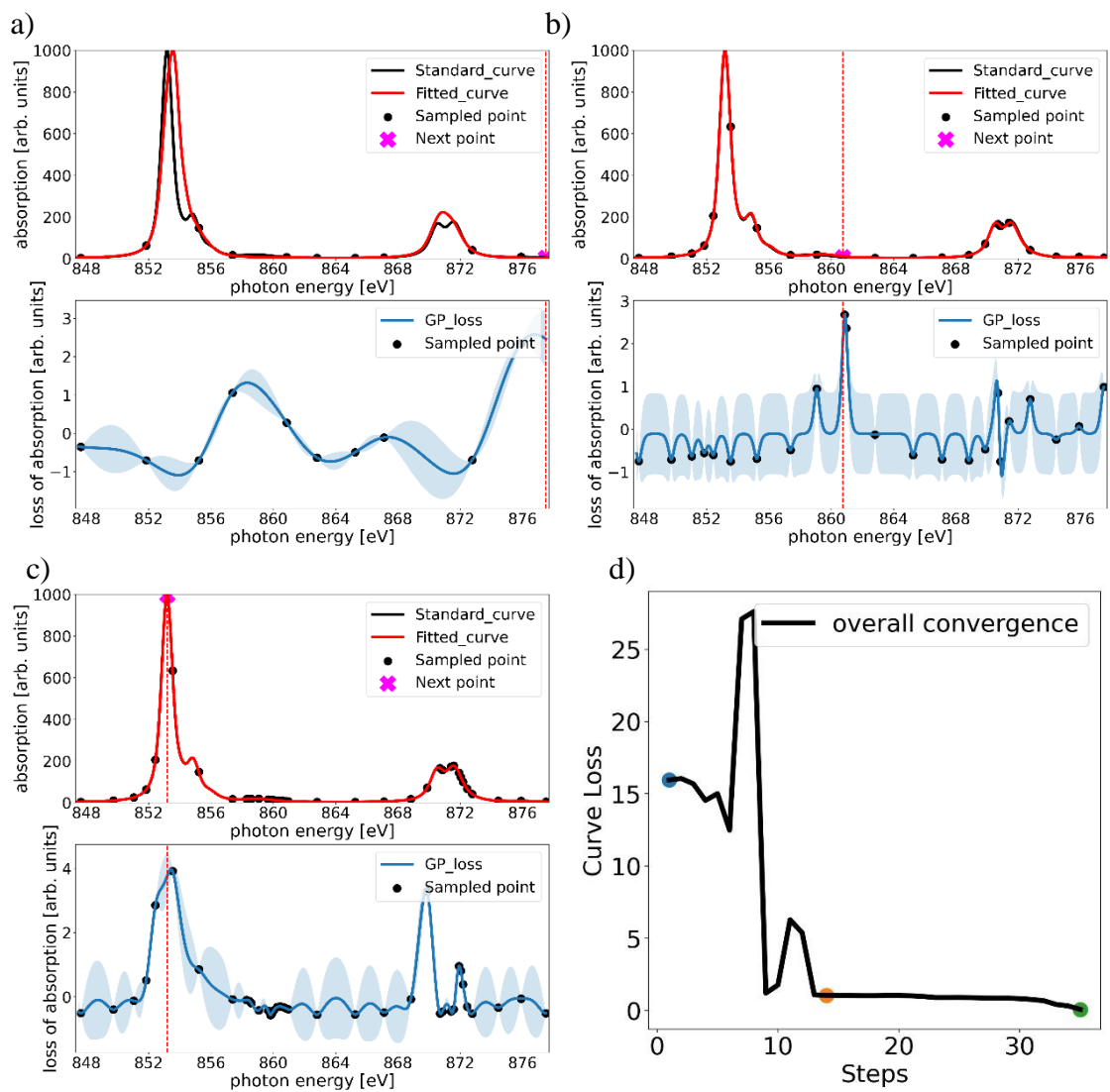

Figure 3. Same as Figure 2, the fitting results of CTM model with: a) 10 sampling points (Step 1); b) 23 sampling points (Step 14); c) Final fitting results using 44 sampling points (Step 35); d) CTM XAS curve convergence, with the colored dots denote in order the results of Figure 3.a to 3.c. The parameter comparison for listed steps is shown in Supplementary Table S6.

**Automatic model selection**

An interesting question is whether an appropriate atomic model can be automatically identified, as recently demonstrated on model selection and information fusion[41] and hypothesis learning[42]. To verify this, we integrate both CFM and CTM models starting with a theoretical standard real spectrum obtained based on either CFM or CTM, and let the ABO algorithm to automatically determine which model should be used. Since the CTM model includes all parameters of CFM model, which means it in theory can fully recover the theoretical standard real spectrum of the CFM by setting the common parameters the same and keeping the rest parameters of ligand field to be 0. So, in our test, in order to allow the two models to be distinguished from each other, we applied different peak broadening methods on different models to ensure that the final generated XAS spectrum have different shapes. Then to avoid the initial arbitrariness, 5 random points are used to train preliminary models. The follow-up convergence behaviors are shown in Figure 4.a and 4.b, for CFM and CTM theoretical standard real spectrum, respectively. Obviously, there is already a clear distinguishment for the CFM and CTM models within 22 iterations, i.e., 22 more sampling points, as measured by the deviations between Loss_CFM and Loss_CTM. That is, the ABO algorithm can easily distinguish the model Hamiltonian and select the correct one based on the experimental data. For instance, in Figure 4.a with a CFM theoretical standard real spectrum, the ABO algorithm using a CTM Hamiltonian can have comparable performance as CFM for less than 7 sampling points (i.e., up to two steps plus the initial five random points). As the number of iterations keeps increasing, the results obtained using CTM Hamiltonian deviate more from the theoretical standard real spectrum in the beginning and then decrease to a constant level. Therefore, we suspect that by training several models in parallel, the important physics for a material system can be automatically identified, e.g., whether the charge transfer processes play an important role. It is noted that the model selection can be performed in a more reliable way by evaluating the Bayes factors[43], which will be saved for future study.

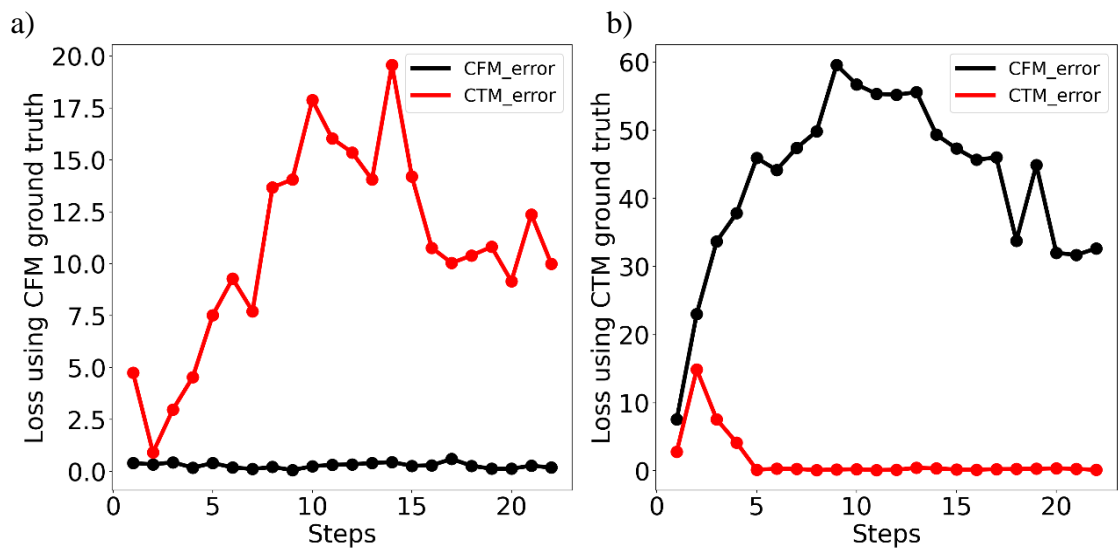

Figure 4. The Loss comparison of CFM and CTM model fBO fitting using different theoretical standard real spectrum: a) CFM theoretical standard real spectrum; b) CTM theoretical standard real spectrum.

**Experimental curve fitting based on the CTM model**

After demonstrating that the ABO algorithm works on the theoretically generated standard real XAS, we turn now to find out its performance on the experimental data. According to Ref.[9], it is clear that the charge transfer processes are critical for NiO, thus CTM is more appropriate to describe the underlying physics of NiO and we consider only CTM here. Starting from the experimental XAS results[40], we firstly fit the CTM model parameters with 12 random sampling points in order to avoid local minima and possible sampling bias. The results of ABO fitting are shown in Figure 5. The comparison between all the parameters is shown in Supplementary section S3, Fig. S5 and the convergence plots of case starts with only one initial points is shown in Fig. S6.

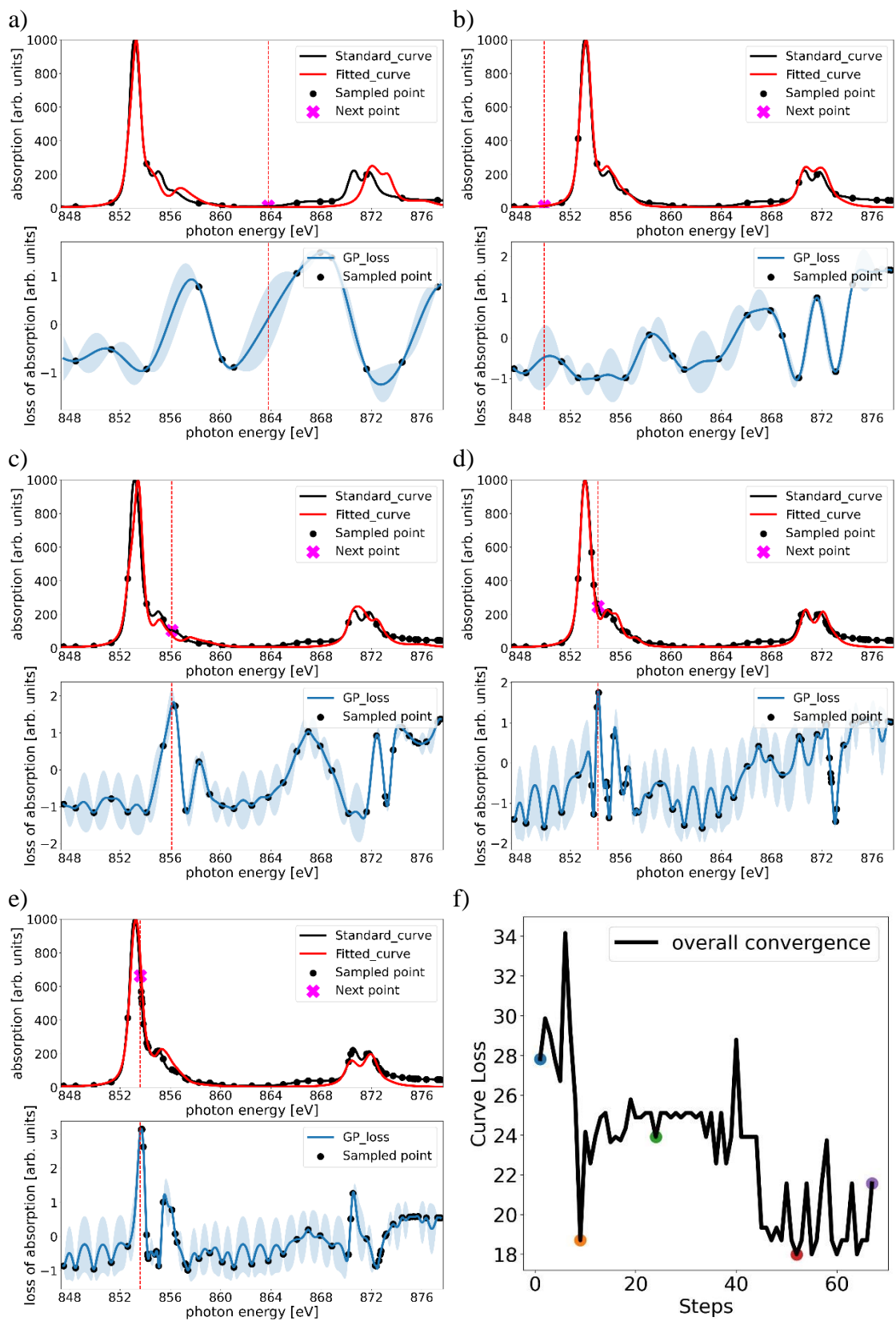

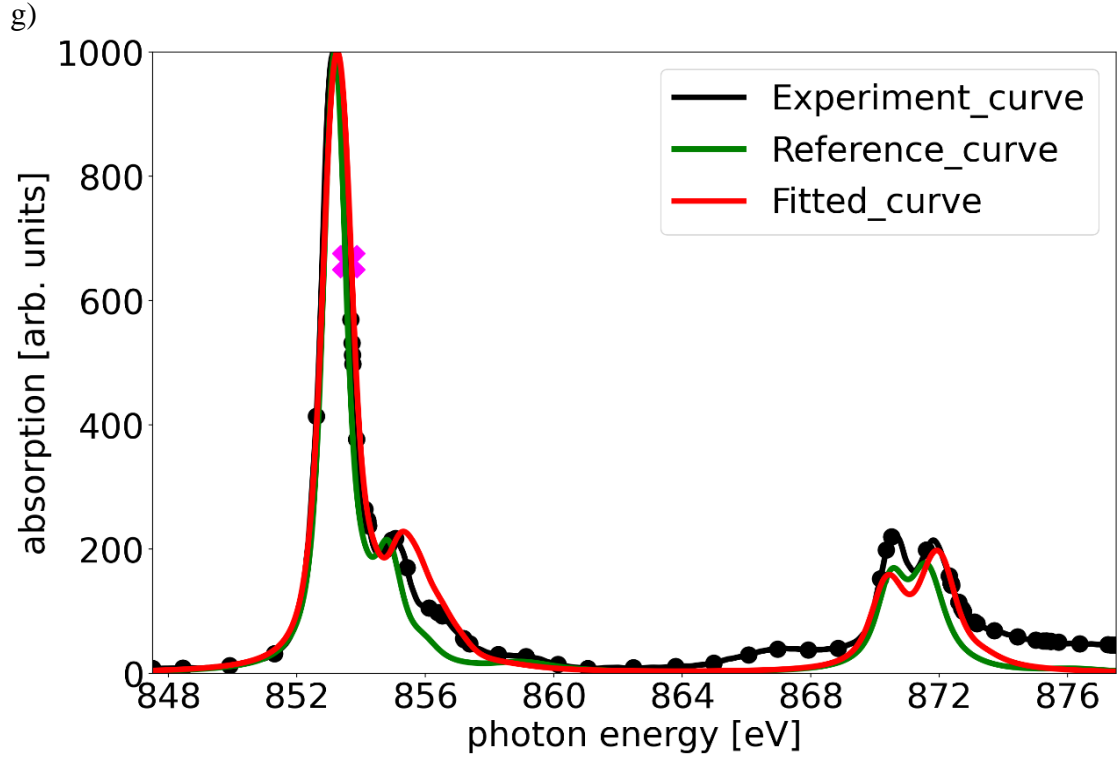

Figure 5. The fitting results of experimental results of: a) 13 sampling points (Step 1); b) 21 sampling points (Step 9); c) 36 sampling points (Step 24); d) 64 sampling points (Step 52); e) 79 sampling points (Step 67); f) XAS curve convergence with respect of experimental curve, with the colored dots denote in order the results of Figure 5.a to 5.e; g) the XAS curve comparison of experimental curve, reference curve and ABO fitted curve.

As shown in Figure 5.a, because of the relatively large initial sample set, the shape of fitted curve is already closer to the experimental XAS, in particular for the $L_2$ edge but not the $L_3$ and satellite peaks. Figure 5.b to Figure 5.e depict the follow-up optimization snapshots. Between Figure 5.b and 5.c, it is observed that the sampling points are clustered between 872 and 876 eV, trying to improve the fitting for the $L_2$ edge. This suggests our ABO algorithm can focus on the most problematic points where the physical model can get most potentially improved. In particular, after a few attempts without improvements, the region between 872 and 876 eV is abandoned and the ABO algorithm begins to explore the $L_3$ peak regions centered at 853.3eV, as clearly marked by the suggested sampling points in Figure 5.c, 5.d, and 5.e. This can be attributed to the fact that the covariance for sampling points between 872 and 876 eV becomes smaller due to the accumulation of points in the region, though the loss in this region is still large, the ABO algorithm prefers to explore other regions that can help improve the results, *e.g.*, the peak regions where the deviations are still large. Therefore, the ABO algorithm can progressively optimize the results on its own if a sufficient number of iterations is allowed, as evidenced by the step descending convergence behavior around the 40-th step (Figure 5.f).

Although a perfect agreement between the model-derived XAS and experimental standard real spectrum cannot be achieved, detailed analysis reveals that the ABO algorithm can be applied to construct a reasonable Hamiltonian by active-learning sampling the experimental XAS. Unlike the previous convergence plots using the curves generated by CFM and CTM as the standard real spectrum where the loss drops rapidly, for the experimental standard real spectrum, the loss function converges slowly and eventually shows an oscillating behavior after 40 iterations (Figure 5.f and supplementary Figure S5), indicating there are more than one solution which cannot be distinguished by varying current set of parameters. We identify three models with distinct parameters characterizing the oscillating region, with the resulting parameters summarized in Table S7. Correspondingly, it is observed that there are four most diverged parameters (i.e., $U_{dd}$, $U_{pd}$, and $F_{dd}^2$, $F_{dd}^4$) among such models (the detailed convergence behaviors of these parameters are shown in supplementary Figure S5). The Slater integral $F_{dd}^2$ and $F_{dd}^4$ are the integral over the radial wave functions in the electron-electron interaction Hamiltonian (the details can be referred to the supplementary S4) and can be calculated in the Hartree-Fock approximation on a free ion using Cowan's code[44]. Physically, $F_{dd}^2 > F_{dd}^4$ and a good approximation gives a constant ratio $F_{dd}^4 = 0.62\ F_{dd}^2$ for $3d^8$ configuration [40]. For the spherical part of the Coulomb repulsion parameters $U_{dd}$ and $U_{pd}$, their values were obtained via fitting the multiplet ligand field model to experimental XAS spectra directly in Ref.[9], which suggest that the parameters obtained by fitting the experimental results using ABO should be comparable with the reference values. However, this is not the case for some fitted parameters in the oscillation zone regarding not only $U_{dd}$ and $U_{pd}$, but also $F_{dd}^2$ and $F_{dd}^4$, thus these possibilities in the oscillation zone are ruled out. Here, we would also like to mention that due to the limitation of the adopted CTM model, it is not surprising that the parameter sets obtained by fitting the spectral shape using ABO do not always demonstrate proper physical meaning.

Based on the loss and physical reasonableness of the fitting parameters, the final result is shown in Figure 5.g. Obviously, XAS obtained from the CTM Hamiltonian fitted using our ABO algorithm exhibits good agreement with the experimental data, in good comparison with XAS obtained using the model parameters obtained by experts[9]. Table 1 summarizes the resulting parameters from Ref.[9] and the final parameters using our ABO algorithm. For Coulomb interaction parameters $G_{pd}^3$, crystal field splitting $10D_q$, hopping integral of $e_g$ symmetry $V_{e_g}$, and exchange magnetic field $B_{ex}$, ABO results show relative strong deviations from the results in the literature. The reason for such deviations can be attributed to the FI of such parameters as shown in Figure S2. This suggests that there is a big degree of freedom in developing physical understanding of XAS, as the same experimental data can be fitted by several groups of atomic Hamiltonian parameters, entailing more detailed theoretical calculations.

Table 1. The CTM model parameters comparison and relative XAS curve loss

| Name | $U_{dd}$ | $U_{pd}$ | $\Delta$ | $F_{dd}^2$ | $F_{dd}^4$ | $F_{pd}^2$ | $G_{pd}^1$ | $G_{pd}^3$ |
|---|---|---|---|---|---|---|---|---|
| Ref.[9] | 7.30 | 8.50 | 4.70 | 11.14 | 6.87 | 6.67 | 4.92 | 2.80 |
| ABO | 8.05 | 10.80 | 4.84 | 10.74 | 8.53 | 8.25 | 5.83 | 1.81 |
| Name | $D_q$ | $D_{qL}$ | $V_{e_g}$ | $V_{t2g}$ | $\xi_{3d}$ | $\xi_{2p}$ | $B_{ex}$ | loss |
| Ref.[9] | 0.56 | 1.44 | 2.06 | 1.21 | 0.081 | 11.51 | 0.12 | 26.96 |
| ABO | 1.60 | 1.20 | 0.49 | 1.22 | 0.120 | 11.35 | 0.29 | 21.57 |

**Experimental curve fitting based on the CTM model with background subtraction and model constrains**

Although our ABO fitted spectra are in good agreement with the experimental curve, the fitted parameter sets still shows some non-physical relations compared with the theoretical value, e.g., Veg<Vt2g. This can be attributed to the following three reasons: 1. The calculated spectrum using the CTM model does not include the continuum absorption background, as the background information in the experimental spectrum is not completely removed; 2. The algorithm sees only the local information, i.e., only the selected-out measurement points rather than the entire curve. The aggregation of samples caused by the non-negligible background can lead to the overweight of regions with concentrated sampling points and cause bias during the parameter fitting; 3. Greedy algorithms without physical constraints only focus on reducing the curve differences, which may give rise to overfitting and thus wrong physical relations.

Correspondingly, we updated our model by restricting $V_{eg} > V_{t2g}$, and eliminated the step edge background by fitting a two-step arctangent weighted function [46–48] after each fBO process, and used the background-eliminated data for the sampling of sBO process. The two-step arctangent weighted function we used is:

$$Background(x) = A_1 \left\{ 0.5 + \frac{\arctan\left[\frac{x - u_1}{c_1}\right]}{\pi} \right\} + A_2 \left\{ 0.5 + \frac{\arctan\left[\frac{x - u_2}{c_2}\right]}{\pi} \right\}$$

where $A$ is the step height, $x$ is the X-ray energy, $u$ is the center of function at the desired X-ray energy and $c$ is the constant controls the slope of the step. All such six parameters $A_1, A_2, u_1, u_2, c_1, c_2$ are fitted automatically in our Bayesian optimization loop.

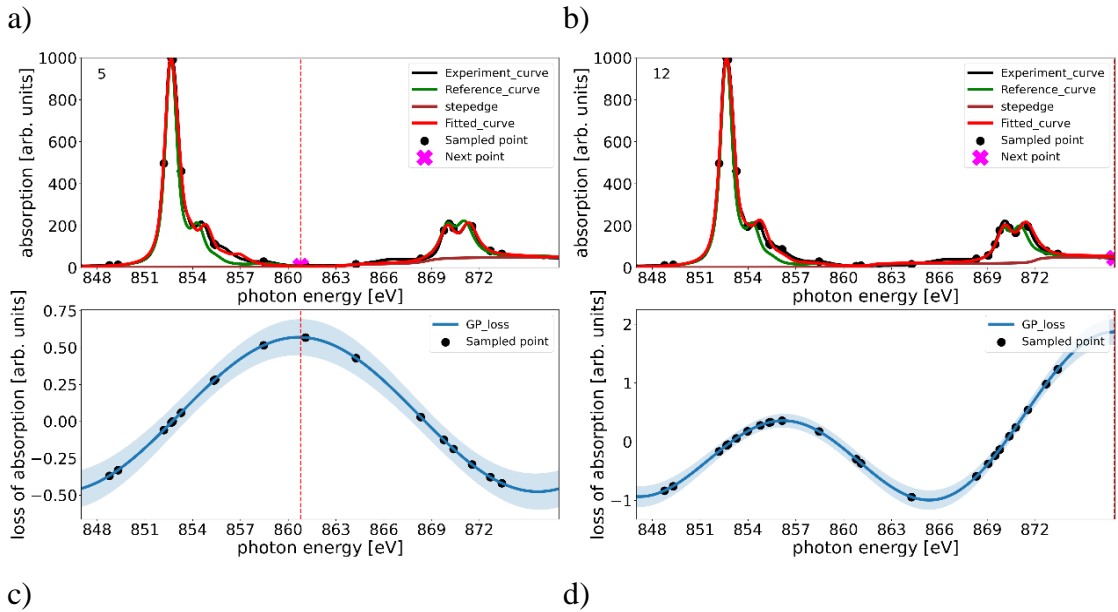

a) b) c) d)

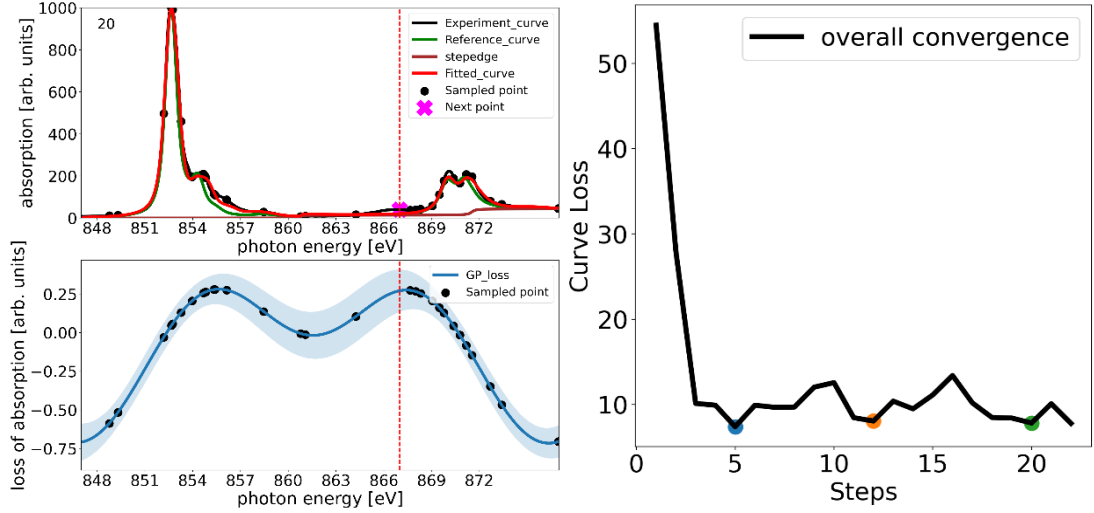

Figure 6. The fitting results of experimental results of: a) 17 sampling points (Step 5); b) 24 sampling points (Step 12); c) 32 sampling points (Step 20); d) XAS curve convergence with respect of experimental curve, with the colored dots denote in order the results of Figure 6.a, 6.b and 6.c.

Table 2. The fitted parameters of CTM model and arctangent weighted function

|  | Ref.[49] | 5 | 12 | 20 |
|---|---|---|---|---|
| $U_{dd}$ | 7.3 | 7.36 | 7.67 | 7.31 |
| $U_{pd}$ | 8.5 | 8.00 | 8.01 | 8.70 |
| $\Delta$ | 4.7 | 3.63 | 3.84 | 4.33 |
| $F^2_{dd}$ | 11.14 | 11.20 | 11.84 | 11.54 |
| $F^4_{dd}$ | 6.87 | 6.84 | 6.57 | 6.78 |
| $F^2_{pd}$ | 6.67 | 6.64 | 6.19 | 6.47 |
| $G^1_{pd}$ | 4.92 | 5.94 | 5.82 | 6.03 |
| $G^3_{pd}$ | 2.8 | 3.58 | 3.07 | 1.84 |
| $10D_q$ | 0.56 | 1.18 | 1.20 | 1.36 |
| $10D_{qL}$ | 1.44 | 2.09 | 1.92 | 1.42 |
| $V_{eg}$ | 2.06 | 1.21 | 1.58 | 1.80 |
| $V_{t2g}$ | 1.21 | 0.83 | 1.12 | 0.88 |
| $\xi_{3d}$ | 0.081 | 0.17 | 0.12 | 0.15 |
| $\xi_{2p}$ | 11.51 | 11.43 | 11.54 | 11.49 |
| $B_{ex}$ | 0.12 | 0.14 | 0.13 | 0.16 |
| $A_1$ |  | 33.57 | 31.32 | 30.41 |
| $A_2$ |  | 16.61 | 19.53 | 13.57 |
| $u_1$ |  | 11.74 | 15.00 | 14.52 |
| $u_2$ |  | 6.69 | 4.75 | 3.81 |
| $c_1$ |  | 0.69 | 0.34 | 0.18 |
| $c_2$ |  | 0.77 | 0.46 | 0.06 |
| *loss* |  | 7.40 | 8.07 | 8.63 |
| *points* |  | 17 | 24 | 32 |

As shown in Figure 6, compared to the previous results without removing the background, the loss (cf. Figure 6(d)) in this fitted curve drops rapidly to below 12 within four iterations (i.e., with 16 sampling points) and stabilized at around 10 in subsequent fits with no significant fluctuations. We find that the sampling points are more distributed throughput the whole energy range (Figure 6.a, 6.b and 6.c), with a tendency to accumulate in regions with fine features in the XAS spectra such as satellite peaks. This demonstrates that our ABO algorithm can explore a wide region while maintaining a high exploitation rate. And for the parameter comparison as shown Table 2, the model fitted more physically reasonable parameters in fewer cycles and effectively identified stable and robust backgrounds during the experiment. The parameters with relative larger deviations this time are $10D_q$ and $10D_{qL}$. The reason of such deviations can be attributed again to the relative lower FI of such parameters as shown in Figure S2 for $10D_q$ and $10D_{qL}$. From the physical point of view, for the deviation of $10D_q$, when measuring the L-edges in XAS, the 2p electrons are mainly excited to the $e_g$ orbitals because the $t_{2g}$ orbitals are fully occupied by the 8 value electrons of $Ni^{2+}$ ions. Therefore, the magnitude of $10D_q$ is not expected to significantly affect the spectral shape, but rather that the global shift of the energy positions corresponding to the absorption peaks and the relative heights between the different peaks. As the calculated XAS spectra are shifted during the ABO process in order to fit the experiment data, only the relative heights remain as a relatively weak condition to obtain 10Dq, which can be also influenced by the values of other parameters. Last but not least, the constraints can be easily derived from DFT calculations, which will be addressed systematically in the future study. We also applied the ABO algorithm to the other material systems such as MnO and $SrTiO_3$ (not shown), and observed that accurate curve fitting can be achieved, in particular with the physical constraints derived from DFT calculations.

**Discussion**

Although the prediction of XAS appears to be a simple one-dimensional regression as a function of energy, it is actually a sophisticated problem which entails physics-driven modelling. As it is demonstrated, by incorporating the physical Hamiltonian model into the fitting process, our ABO algorithm can not only predict the peak/satellite positions and fine structures of XAS, but also can automatically select and inversely construct the physical model simultaneously in an adversarial manner. Importantly, we find that the sampling efficiency using our ABO algorithm can be significantly enhanced in comparison to a stopping criterion algorithm based on Bayesian optimization as done in Ref.[23]. For instance, for the CFM model theoretical standard real spectrum, the number of sampling points based on our ABO algorithm is 24.3% of that for a single BO with stopping criterion 0.025, and the corresponding ratio for the CTM model theoretical standard real spectrum is 18.8% (cf. Supplementary Table S8). That is, the physics-driven Hamiltonian obtained via the fBO component is essential for more effective sampling of XAS. Furthermore, the high accuracy and efficiency of our ABO algorithm can be attributed to two reasons. The first reason is that when fBO and sBO compete against each other on loss, the algorithm is actually trying to drive the sample distribution closer to the true data distribution. Such a distribution not only effectively represents the important information of the data, but also allows for a more efficient and faster fitting of the model. The second reason is that the introduction of physical models within fBO changes the prior of sBO in each iteration, this makes it possible to explore extensively while maintaining a high exploitation ratio, *i.e.*, the high exploration rate of the fBO provides us with a wide range of possible model spectra, which leads to variations in losses between measurements and physical models, while sBO with high exploitation ratio can precisely pinpoint the maximum losses during the sampling, and thus it is able to capture the peak/satellite positions of the XAS spectra efficiently and accurately. However, it should not be overlooked that such efficient sampling is based on regression analysis of the current data using a physical model, a process that prolongs the decision-making time in the experimental process. The current version of ABO takes about 15 minutes per round of computation on a single-core NVIDIA Tesla T4 GPU and intel core AVX512. Depending on the complexity of the model, it usually takes between one to five rounds of calculations to find the minimum that satisfies the threshold. This shortcoming on the one hand can be partially overcome by performing batch sampling in sBO or by multi-threaded parallel fitting in fBO; on the other hand, by combining with DFT calculation, automatically pre-determining some of the parameters, the fitting process can be even faster, so that our ABO algorithm can get integrated with experimental measurements in the future. One thing worth to mention is that, besides the application of ABO approach on XAS spectrum, this approach can be easily transferred to other problems with physical properties derived based on parameterized Hamiltonian, which can be correlated with experimental measurements. For example, by employing advanced Fourier basis in Ref. [50] as the physical model, our ABO can be applied to facilitate the fitting and sampling of EXAFS, which will be saved for future investigation.

To summarize, we implemented an ABO algorithm for physics-informed active learning sampling of XAS. Applying the algorithm on the simulation of two different theoretical standard real spectra, it is demonstrated that our algorithm not only succeeds in predicting the true curve, but also accurately predicts the parameters in the atomic Hamiltonian. Intuitive application of the ABO algorithm shows that hypothesis learning can be accomplished, so that the physically meaningful models can be automatically selected. For simulations using real experimental data, our ABO algorithm can quickly find the optimal solution of the model for the current data and select the sampling points most likely to further enhance the model accordingly, and it is able to automatically adjust the XAS spectra background and the model used to better predict the experimental data as it evolves. We believe that the ABO algorithm has a great potential for real-time applications in XAS experiments with the on-the-fly construction of physical models.

## Data availability

All data needed to produce the work are available from the corresponding author.

## Code availability

ABO code is available upon reasonable request from the corresponding author.

## Acknowledgements

The authors appreciate Gabriel Gomez for insightful discussions, and gratefully acknowledge computational time on the Lichtenberg High-Performance Supercomputer. Yixuan Zhang thanks the financial support from the Fulbright-Cottrell Award. This work was also supported by the Deutsche Forschungsgemeinschaft (DFG, German Research Foundation) – Project-ID 405553726 – TRR 270. We also acknowledge support by the Deutsche Forschungsgemeinschaft (DFG – German Research Foundation) and the Open Access Publishing Fund of Technical University of Darmstadt.

## Author information

**Affiliations**

**Institute of Materials Science, Technical University of Darmstadt, Darmstadt 64287, Germany**
Yixuan Zhang, Ruiwen Xie, Teng Long, Hongbin Zhang.
**School of Materials Science and Engineering, Shandong University, Jinan 250061, China**
Teng Long

**Faculty of Physics, University of Duisburg-Essen and Center for Nanointegration Duisburg-Essen (CENIDE), Duisburg 47057, Germany**
Damian Günzing, Heiko Wende, Katharina J. Ollefs.

## Contributions

This work originated from the discussion of H.Z., Y.Z., R.X and T.L. H.Z. and R.X. supervised the research. Y.Z. and T.L. worked on the machine learning model. Y.Z. and R.X. worked on the Quanty simulations. Y.Z., R.X., D.G., H.W. and K.O. worked on data analysis. All authors contributed in the writings.

## Corresponding authors

Correspondence to Hongbin Zhang.

## Ethics declarations

## Competing interests

The authors declare no competing interests.